\documentclass[aps,prb,twocolumn,showpacs]{revtex4}

\usepackage{graphicx}
\def  \bsig    {\mbox{\boldmath$\sigma$}}

\begin{document}

\title{Reflection of electrons from a domain wall in magnetic nanojunctions}

\author{V. K. Dugaev$^{1,2}$, J. Berakdar$^1$, and J. Barna\'s$^{3,4}$}
\affiliation{$^1$Max-Planck-Institut f\"ur Mikrostrukturphysik, Weinberg 2, 06120 Halle, Germany}
\affiliation{$^2$Institute for Problems of Materials Science, National Academy of Sciences of Ukraine,
Vilde 5, 58001 Chernovtsy, Ukraine}
\affiliation{$^3$Department of Physics, Adam Mickiewicz University, Umultowska 85, 61-614~Pozna\'n, Poland}
\affiliation{$^4$Institute of Molecular Physics, Polish Academy of Sciences,
M.~Smoluchowskiego~17, 60-179~Pozna\'n, Poland}

\date{\today }

\begin{abstract}
\vskip0.2cm \noindent Electronic transport through thin and
laterally constrained domain walls in ferromagnetic nanojunctions
is analyzed theoretically. The description is formulated in the
basis of scattering states. The resistance of the domain wall is
calculated in the regime of strong electron reflection from the
wall. It is shown that the corresponding magnetoresistance can be
large, which is in a qualitative agreement with recent
experimental observations. We also calculate the spin current
flowing through the wall and the spin polarization of electron gas
due to reflections from the domain wall.
\end{abstract}

\pacs{75.60.Ch,75.70.Cn,75.70.Pa}

\maketitle

\section{Introduction}

There is a growing interest in the resistance and
magnetoresistance associated with domain walls (DWs) in metallic
ferromagnets.\cite{kent01} Owing to recent progress in
nanotechnology, it became possible now to extract a single DW
contribution to electrical
resistance.\cite{kent01,hong98,rudiger98,kent99} Surprisingly, it
turned out that the resistance of a system with DWs in some cases
was smaller than in the absence of DWs,\cite{hong98,rudiger98}
whereas in other cases it was
larger.\cite{gregg96,garcia99,ebels00} This intriguing
observation led to considerable theoretical interest in
electronic transport through
DWs.\cite{tatara97,levy97,gorkom99,jonkers99,tatara01} The
interest is additionally stimulated by possible applications of
the associated magnetoresistance in magnetoelectronics devices.

In a series of experiments the magnetoresistance associated with
DWs was found to be very
large.\cite{gregg96,garcia99,ebels00,danneau02} Moreover, recent
experiments on Ni microjunctions showed that constrained DWs at
the contact between ferromagnetic wires produce an unexpectedly
large contribution to electrical resistivity, and consequently
lead to a huge negative magnetoresistance.\cite{chopra02} It was
shown theoretically\cite{bruno99} that DWs in magnetic
microjunctions can be very sharp, with the characteristic width
$L$ being of atomic scale. This is much less than typical DW
width in bulk materials or thin films.

Theoretical descriptions of the transport properties of DWs are
mainly restricted to very smooth DW,
\cite{levy97,gorkom99,cabrera74,brataas99,simanek01,dugaev02}
which is more appropriate for bulk ferromagnets. Electron
scattering from DWs is then rather small and the spin of an
electron propagating across the wall follows the magnetization
direction almost adiabatically. The additional resistance
calculated in the semiclassical approximation can be either
positive or negative (depending on material parameters) and
rather small. The validity condition for the semiclassical
approximation is $k_{F\uparrow (\downarrow )}L\gg 1$, where
$k_{F\uparrow }$ and $k_{F\downarrow}$ are the Fermi wavevectors
for the majority and minority electrons, respectively.

For sharp DWs, however, scattering of electrons from the wall is
significant and the semiclassical approximation is no longer
applicable. Some numerical calculations of the magnetoresistance
in magnetic nanojunctions have been presented in
Ref.~[\onlinecite{imamura00}] in the context of the conductance
quantization in microjunctions due to lateral confinement. The
problem of large magnetoresistance in magnetic junctions was also
analyzed recently by Tagirov {\it et al},\cite{tagirov} where DW
was approximated by a potential barrier independent of the
electron spin orientation. The ballistic regime of electron
transport through the domain wall has been also considered using
some numerical simulations\cite{hoof99} and {\it ab initio}
calculations.\cite{kudrnovsky00,kudrnovsky01,yavorsky02}

In this paper we consider the case of a  thin DW, when the
condition $k_{F\uparrow (\downarrow )}L\lesssim 1$ is fulfilled
(the semiclassical approximation is not applicable). In the limit
of $k_{F\uparrow (\downarrow )}L\ll 1$, we formulate the problem
as a transmission of electrons through a potential barrier. Such a
formulation can be treated analytically. In addition, we restrict
our considerations to the case of DWs with very small lateral
dimensions, when only a single quantum channel takes part in
electronic transport. We show, that the magnetoresistance
associated with DWs can be rather large - up to 70\%, depending on
the polarization of electrons.

In section 2 we describe the model and introduce the basis of
scattering states. Conductance of a domain wall is calculated in
Section 3. Spin current flowing through DW and spin polarization
of the electron gas due to reflections from the wall are
calculated in Sections 4 and 5, respectively. Summary and final
remarks are in Section~6.

\section{Model and scattering states}

Let us consider conduction electrons described by a parabolic
band, which propagate in a spatially nonuniform magnetization
${\bf M}({\bf r})$. The system is then described by the following
Hamiltonian:
\begin{equation}
\label{1}
H=-\frac{\hbar ^2}{2m}\, \frac{\partial ^2}{\partial
{\bf r}^2} -J\bsig \cdot {\bf M}({\bf r}),
\end{equation}
where $J$ is the exchange integral and $\bsig =(\sigma _x,\,
\sigma _y,\, \sigma _z)$ are the Pauli matrices. For a domain
wall with its center localized at $z=0$ we assume ${\bf
M}(z)=\left[ M_0\sin \varphi (z),\, 0,\, M_0\cos \varphi
(z)\right]$, where $\varphi (z)$ varies from zero to $\pi $ for
$z$ changing from $z=-\infty $ to $z=+\infty $. Let the
characteristic length scale of this change be $L$ (refereed to in
the following as the DW width).

When DW is laterally constrained, the number of quantum transport
channels can be reduced to a small number. In the extreme case
only a single conduction channel is active. In such a case, one
can restrict considerations to the corresponding one-dimensional
model, and rewrite the Hamiltonian (1) as
\begin{equation}
\label{2} H=-\frac{\hbar ^2}{2m}\, \frac{d^2}{dz^2}-JM_z(z)\,
\sigma _z-J M_x(z)\, \sigma _x\, .
\end{equation}
Although this model describes only a one-channel quantum wire, it
is sufficient to account qualitatively for some of the recent
observations. Apart from this, it can be rather easily generalized
to the case of a wire with a few conduction channels.

In the following description we use the basis of scattering
states. The asymptotic form of such states (taken sufficiently
far from DW) can be written as
\begin{eqnarray}
\label{3}
\chi_{R\uparrow }(z)=
\left\{
\begin{array}{cc}
\left(
\begin{array}{c}
e^{ik_\uparrow z}+r_{R\uparrow }\, e^{-ik_\uparrow z} \\
r_{R\uparrow }^f\, e^{-ik_\downarrow z}
\end{array}
\right) , & z\ll -L, \\
\\
\left(
\begin{array}{c}
t_{R\uparrow }\, e^{ik_\downarrow z} \\
t_{R\uparrow }^f\, e^{ik_\uparrow z}
\end{array}
\right) , & z\gg L,
\end{array}
\right.
\end{eqnarray}
where $k_{\uparrow (\downarrow )}=\left[ 2m(E\pm  M)\right] ^{1/2}/\hbar $,
with $M=JM_0$ and $E$ denoting the electron energy. The scattering
state (3) describes the electron wave in the spin majority channel
incident from $z=-\infty $, which is partially reflected into the
spin-majority and spin-minority channels, and also partially
transmitted into these two channels. The coefficients
$t_{R\uparrow }$ and $t_{R\uparrow }^f$ are the transmission
amplitudes without and with spin reversal, respectively, whereas
$r_{R\uparrow }$ and $r_{R\uparrow }^f$ are the relevant
reflection amplitudes. It is worth to note that transmission from
the spin-majority channel at $z<0$ to the spin-majority channel at
$z>0$ requires spin reversal. The scattering states corresponding
to the electron wave incident from $z=-\infty $ in the
spin-minority channel have a similar form. Also similar form have
the scattering states describing electron waves incident from the
right to left.

In a general case the transmission and reflection coefficients
are calculated numerically, as described in the next section.
When $k_{F\uparrow (\downarrow )}L\ll 1$, then the coefficients
can be calculated analytically. Upon integrating the
Schr\"odinger equation $H\psi =E\psi $ (with the Hamiltonian
given by Eq.~(2)) from $z=-\,\delta $ to $z=+\,\delta $, and
assuming $L\ll \delta \ll k_{\uparrow (\downarrow )}^{-1}$, one
obtains
\begin{equation}
\label{4}
-\frac{\hbar }{2m}\left( \left. \frac{d\chi }{dz}\right|
_{z=+\delta } -\left. \frac{d\chi }{dz}\right| _{z=-\delta
}\right) -\lambda\,\sigma _x \, \chi (z=0)=0
\end{equation}
for each of the scattering states (for clarity of notation the
index of the scattering states is omitted here), where
\begin{equation}
\label{5}
\lambda\simeq \frac{J}{\hbar }\int _{-\infty }^\infty dz\; M_x(z).
\end{equation}
Equation (4) has the form of a spin-dependent condition for
electron transmission through a $\delta $-like potential barrier
located at $z=0$. To obtain this equation we also used the
condition $k_{\uparrow (\downarrow )}L\ll 1$, which is opposite to
the condition used in the semiclassical approximation. The
magnitude of the parameter $\lambda$ in Eq.~(5) can be estimated
as $\lambda\simeq JM_0L/\hbar =ML/\hbar $.

Using the full set of scattering states and the condition (4),
together with the wave function continuity condition, one finds
the transmission amplitudes
\begin{eqnarray}
\label{6}
t_{R\uparrow (\downarrow )} =t_{L\downarrow (\uparrow )}
=\frac{2v_{\uparrow (\downarrow )}(v_\uparrow +v_\downarrow )}
{(v_\uparrow +v_\downarrow)^2+4\lambda^2}\, ,
\nonumber \\
t_{R\uparrow (\downarrow )}^f =t_{L\downarrow (\uparrow )}^f
=\frac{4i\lambda\, v_{\uparrow (\downarrow )}} {(v_\uparrow
+v_\downarrow)^2+4\lambda^2}\, ,
\end{eqnarray}
where $v_{\uparrow (\downarrow )}=\hbar k_{\uparrow (\downarrow )}/m$
denotes the electron velocity in the spin-majority (spin-minority)
channel.

According to Eq.~(6), the magnitude of spin-flip transmission
coefficient can be estimated as (for simplicity we omit here the
state indices)
\begin{equation}
\label{7} \left| t^f\right|^2 \sim \left(\frac{\lambda
v}{v^2+\lambda^2}\right)^2 \sim \left(\frac{ M\varepsilon
_0}{\varepsilon _F\, \varepsilon _0+ M^2}\right)^2\; (k_FL)^2,
\end{equation}
where $\varepsilon _F=\hbar ^2k_F^2/2m$ and $\varepsilon _0=\hbar ^2/mL^2$. For
$k_FL\ll 1$ one finds $\varepsilon _0\gg \varepsilon _F$. Thus,
taking $\varepsilon _F\sim M$, one obtains
\begin{equation}
\label{8} \left| t^f\right|^2 \sim \left(\frac{M}{\varepsilon
_F}\; k_FL\right)^2\ll 1.
\end{equation}
Accordingly, a sharp domain wall can be considered as an
effective barrier for the spin-flip transmission. On the other
hand, the probability of spin conserving transmission is much
larger, $\left| t/t^f\right|^2 \sim  \varepsilon _F \varepsilon
_0/ M^2\gg 1$. This means that electron spin does not follow
adiabatically the magnetization direction when it propagates
through the wall, but its orientation is rather fixed.

It is worth to note, that the conservation of flow in the
spin-dependent case considered here has the following form
\begin{equation}
\label{9}
v_\uparrow \left( 1-\left| r_{R\uparrow }\right| ^2\right)
-v_\downarrow \left| r_{R\uparrow }^f\right| ^2
=v_\downarrow \left| t_{R\uparrow }\right| ^2
+v_\uparrow \left| t_{R\uparrow }^f\right| ^2,
\end{equation}
and also analogous equations for the other scattering states.

\section{Resistance of the domain wall}

To calculate conductance of the system under consideration, let
us start with the current operator
\begin{equation}
\label{10}
\hat{j}(z)=e\, \psi ^\dag (z)\, \hat{v}\, \psi (z),
\end{equation}
where $\hat{v}$ is the velocity operator, whereas $\psi ^\dag
(z)$ and $\psi (z)$ are the electron field operators taken in the
spinor form. Accordingly, the form of Eq.~(10) implies summation
over spin components. Using the expansion of $\psi (z)$ over the
scattering states (3) and carrying out the quantum-mechanical
averaging, one obtains the following formula for the current
\begin{equation}
\label{11}
j(z)=-ie\sum _n\int \frac{dk}{2\pi }\int
\frac{d\varepsilon }{2\pi}\; {\rm e}^{i\varepsilon \eta}\,
G_n(k,\varepsilon )\; \chi _n^\dag (z)\, \hat{v}\,  \chi _n(z),
\end{equation}
where $n$ is the index of scattering states ($n={R\uparrow },\,
{R\downarrow },\, {L\uparrow }$, and $L\downarrow $) and $\eta =0^+$.
The matrix elements of the velocity operator $\hat{v}=-(i\hbar /m)\,
\partial /\partial z$ in the basis of scattering
states have the form
\begin{eqnarray}
\label{12}
v_{R\uparrow}\equiv \left< R\uparrow \left| \,
\hat{v}\, \right| R\uparrow \right> =v_\downarrow \left|
t_{R\uparrow }\right| ^2 +v_\uparrow \left| t_{R\uparrow
}^f\right| ^2,
\nonumber \\
v_{R\downarrow}\equiv \left< R\downarrow \left| \, \hat{v}\,
\right| R\downarrow \right> =v_\downarrow \left| t_{R\downarrow
}^f\right| ^2 +v_\uparrow \left| t_{R\downarrow }\right| ^2,
\nonumber \\
v_{L\uparrow}\equiv \left< L\uparrow \left| \, \hat{v}\, \right|
L\uparrow \right> =-v_\uparrow \left| t_{L\uparrow }\right| ^2
-v_\downarrow \left| t_{L\uparrow }^f\right| ^2,
\nonumber \\
v_{L\downarrow}\equiv \left< L\downarrow \left| \, \hat{v}\,
\right| L\downarrow \right> =-v_\uparrow \left| t_{L\downarrow
}^f\right| ^2 -v_\downarrow \left| t_{L\downarrow }\right| ^2.
\end{eqnarray}
Finally, the retarded Green function $G_n(k,\varepsilon )$ in
Eq.~(11) is diagonal in the basis of scattering states.

When the transmission of electrons through the barrier is small,
one can assume that the chemical potential drops at the wall and
is constant elsewhere, $\mu =\mu _R$ for $z<0$ and $\mu =\mu _L$
for $z>0$. This corresponds to the voltage drop $U=(\mu _R-\mu
_L)/e$ across the domain wall, whereas the resistance of the wire
parts outside the wall can be neglected. The Green function
$G_{R\uparrow }(k,\varepsilon )$ acquires then the following
simple form
\begin{equation}
\label{13}
G_{R\uparrow}(k,\varepsilon) =\frac1{\varepsilon
-\varepsilon _{R\uparrow}(k)+\mu _R+i\eta }\, ,
\end{equation}
where $\varepsilon _{R\uparrow}(k)=\hbar ^2k^2/2m- M$. The other
components of the Green function have a similar form.

After integrating over $\varepsilon $ in Eq.~(11) we obtain
\begin{eqnarray}
\label{14}
j(z)=e\int \frac{dk}{2\pi }\left\{
\chi _{R\uparrow }^\dag (z)\, \hat{v}\, \chi _{R\uparrow }(z)\;
\theta \left[ \mu _R-\varepsilon _{R\uparrow }(k)\right]
\right. \nonumber \\ \left.
+\chi _{R\downarrow }^\dag (z)\, \hat{v}\,  \chi _{R\downarrow }(z)\;
\theta \left[ \mu _R-\varepsilon _{R\downarrow }(k)\right]
\right. \nonumber \\ \left.
+\chi _{L\uparrow }^\dag (z)\, \hat{v}\,  \chi _{L\uparrow }(z)\;
\theta \left[ \mu _L-\varepsilon _{L\uparrow }(k)\right]
\right. \nonumber \\ \left.
+\chi _{L\downarrow }^\dag (z)\, \hat{v}\, \chi _{L\downarrow }(z)\;
\theta \left[ \mu _L-\varepsilon _{L\downarrow }(k)\right] \right\} .
\end{eqnarray}
Since the current does not depend on $z$ due to the charge
conservation law, it can be calculated at arbitrary point, say at
$z=0$. Apart from this, the contribution from the states with
$\varepsilon _{R\uparrow (\downarrow )}(k),\, \varepsilon
_{L\uparrow (\downarrow )}(k)\leq \min (\mu _L, \mu _R)$ vanishes
and only the states in the energy range from $\min (\mu _L,\mu
_R)$ to $\max (\mu _L,\mu _R)$ contribute to the current.

\begin{figure}
\includegraphics[scale=0.45]{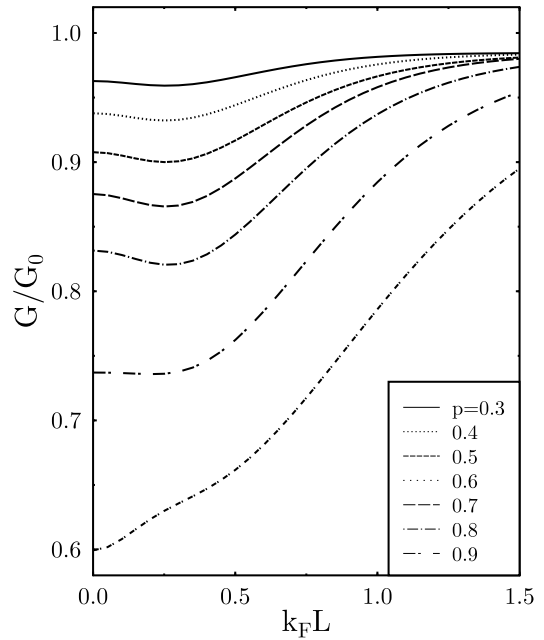}
\caption{Relative conductance of the domain wall as a function of
its width $L$ in a magnetic quantum wire, calculated for
indicated values of the parameter $p\equiv M/\varepsilon_F$.}
\end{figure}

Using Eqs.~(3) and (12) to (14), one obtains the conductance G as
a linear response to small perturbation ($U\rightarrow 0$),
\begin{equation}
\label{15}
G =\frac{e^2}{2\pi \hbar}\left(
\frac{v_\downarrow}{v_\uparrow}\left| t_{R\uparrow }\right| ^2
+\left| t_{R\uparrow }^f\right| ^2
+\frac{v_\uparrow}{v_\downarrow}\left| t_{R\downarrow
}\right| ^2 +\left| t_{R\downarrow }^f\right| ^2\right) ,
\end{equation}
where all the velocities and transmission coefficients are taken
at the Fermi level.

When $k_{F\uparrow (\downarrow )}L\ll 1$, then taking into account
Eq.~(6), one can write the conductance in the form
\begin{equation}
\label{16}
G =\frac{4e^2}{\pi \hbar }\; \frac{v_\uparrow \, v_\downarrow
\left( v_\uparrow +v_\downarrow \right) ^2
+2\lambda^2\left( v_\uparrow ^2+v_\downarrow ^2\right) }
{\left[ \left( v_\uparrow +v_\downarrow \right) ^2
+4\lambda^2\right] ^2}.
\end{equation}
In the limit of $v_\uparrow =v_\downarrow $ and
$\lambda\rightarrow 0$, we obtain the conductance of a
one-channel spin-degenerate wire, $G_0=e^2/\pi \hbar $. In the regime
of ballistic transport $G _0$ is also the conductance of the
investigated system without DW.

Variation of the conductance $G$ with the wall width $L$ (Fig.1)
was calculated from Eq.~(15), with the transmission coefficients
determined numerically. Thus, the results shown in Fig.~1, are
valid for arbitrary value of $k_FL$. The numerical modeling has
been done by direct calculation of the spinor wave function using
Eq.~(2), starting at $z\gg L$ in a form of two transmitted spin up
and down waves with arbitrary numerical coefficients. Then we
restored the function in the region $z\ll -L$ and, by numerical
projecting the obtained spin components on the right- and
left-moving waves (in accordance with Eq.~(3)), we found the
amplitudes of incident and reflected waves.

In the limit $k_FL\ll 1$, the results shown in Fig.1 should
coincide with those obtained from the formula (16). Comparison of
the results obtained from direct numerical calculations and those
obtained from Eq.(16) is shown in Fig.~2. Indeed the results
coincide for $k_FL\ll 1$, whereas at larger values of $k_FL$ the
deviations are large and grow with increasing $k_FL$.

The conductance in the presence of a domain wall is substantially
smaller than in the absence of the wall. Accordingly, the
associated magnetoresistance can be large. For example, for $p$=0.9
in Fig.1 the magnetoresistance is equal to about 70\% (which
corresponds to $G/G_0=0.6$).

\begin{figure}
\includegraphics[scale=0.4]{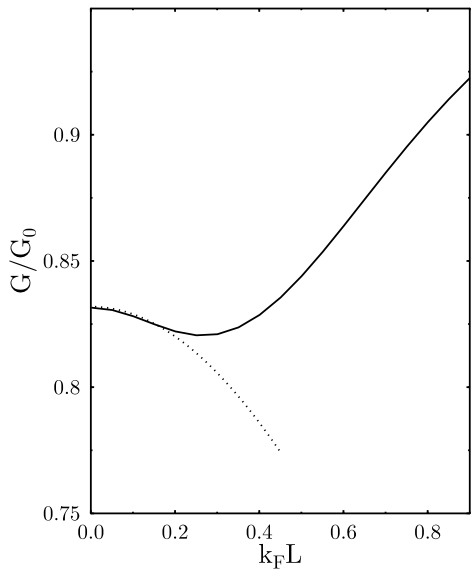}
\caption{Comparison of the results obtained from direct numerical
calculations (solid line) and and from Eq.~(16) valid at $k_FL\ll
1$ (dotted line) for $p=0.7$.}
\end{figure}

It should be noted that in a real magnetoresistance experiment on
magnetic semiconductor nanowires, for which the inequality
$k_FL\gg 1$ can be easily fulfilled, one can have more than one
domain walls. Accordingly, the magnetoresistance effect can be
significantly enhanced.

It is also worth to note that the resistance of an abrupt domain
wall can be smaller than the resistance of a domain wall with
finite (nonzero) thickness. This follows directly from the weak
minimum in some of the curves in Fig.~1 (see also Fig.~2). The
existence of this minimum is related to the sign of the second
derivative of the function $G(\lambda )$ in Eq.~(16), calculated
at $\lambda =0$ (the first derivative vanishes there). In our
simple model, the corresponding sign is negative for
$(v_\downarrow/v_\uparrow ) <(v_\downarrow/v_\uparrow
)_{cr}=2-\sqrt{3}\simeq 0.268$ (or, equivalently, $p<p_c\simeq
0.866$), and positive for $(v_\downarrow/v_\uparrow )
<(v_\downarrow/v_\uparrow )_{cr}$ (i.e., $p>p_c$). When $\lambda $
increases from $\lambda =0$, the conductance decreases in the
former case and increases in the latter one. On the other hand, we
know that for thick domain walls the conductance increases with
increasing wall thickness. Thus, the minimum should occur for the
curves corresponding to $p<p_c$. In the case of strong
polarization, $p>p_c$, the main contribution to the conductance is
associated with the spin-flip transmission through the domain
wall, and the conductance increases monotonously with the width of
the domain wall, in accordance with Eq.~(8).

\section{Spin current}

When the electric current is spin polarized and when there is some
asymmetry between the two spin channels, the flow of charge is
accompanied by a flow of spin (angular momentum). The
$z$-component of the spin current can be calculated from the
following definition of the corresponding spin-current operator
\begin{equation}
\label{17}
\hat{J}_z(z)=\psi ^\dag (z)\, \sigma_z \, \hat{v}\, \psi (z),
\end{equation}
which leads to the following average value
\begin{equation}
\label{18}
J_z(z)=-i\sum _n\int \frac{dk }{2\pi}\int
\frac{d\varepsilon }{2\pi}\; {\rm e}^{i\varepsilon \delta}\,
G_n(k,\varepsilon )\; \chi _n^\dag (z)\, \sigma _z\hat{v}\, \chi
_n(z).
\end{equation}
After carrying out the calculations similar to those described in
the preceding section, one arrives in the linear response regime
(limit of small bias voltage $U$) at the following formulas for
the spin current $J_z$:
\begin{eqnarray}
\label{19}
J_z(z<-L)\hskip7cm
\nonumber \\
=\frac{eU}{2\pi \hbar}\left( \frac{v_\downarrow}{v_\uparrow}\left|
t_{R\uparrow }\right| ^2 +\left| t_{R\uparrow }^f\right| ^2
-\frac{v_\uparrow}{v_\downarrow}\left| t_{R\downarrow }\right| ^2
-\left| t_{R\downarrow }^f\right| ^2\right) ,\hskip0.5cm
\end{eqnarray}
\begin{eqnarray}
\label{20}
J_z(z>L)\hskip7cm
\nonumber \\
=\frac{eU}{2\pi \hbar }\left( \frac{v_\downarrow}{v_\uparrow}\left|
t_{R\uparrow }\right| ^2 -\left| t_{R\uparrow }^f\right| ^2
-\frac{v_\uparrow}{v_\downarrow}\left| t_{R\downarrow }\right| ^2
+\left| t_{R\downarrow }^f\right| ^2\right) .\hskip0.5cm
\end{eqnarray}
Using Eqs.~(6) we find
\begin{equation}
\label{21}
J_z(z>L)=-\frac{8eU}{\pi \hbar}\;
\frac{\lambda ^2\left( v_\uparrow ^2 -v_\downarrow ^2\right) }
{\left[ \left( v_\uparrow +v_\downarrow
\right) ^ 2+4\lambda^2\right] ^2}\;
\end{equation}
and $J_z(z<-L)=-J_z(z>L)$. The magnetic torque due to spin transfer
to the magnetic system within the domain wall is determined by the
non-conserved spin current
\begin{equation}
\label{22}
T(U)=\frac{16eU}{\pi \hbar }\;
\frac{\lambda ^2\left( v_\uparrow ^2 -v_\downarrow ^2\right) }
{\left[ \left( v_\uparrow +v_\downarrow
\right) ^ 2+4\lambda^2\right] ^2}\; .
\end{equation}

It should be noted that spin-flip scattering due to
DW does not allow to separate spin channels like it was in the case
for homogeneous ferromagnets. If we define now the spin conductance
$G_s$ as $G_s=J_z/U$, then one can write for $z>0$
\begin{equation}
\label{23}
G_s=-\frac{8e}{\pi \hbar }\;
\frac{\lambda ^2\left( v_\uparrow ^2 -v_\downarrow ^2\right) }
{\left[ \left( v_\uparrow +v_\downarrow
\right) ^ 2+4\lambda^2\right] ^2}\; .
\end{equation}
Thus, $G_s$ is negative for $z>0$ and positive for $z<0$.

\begin{figure}
\includegraphics[scale=0.45]{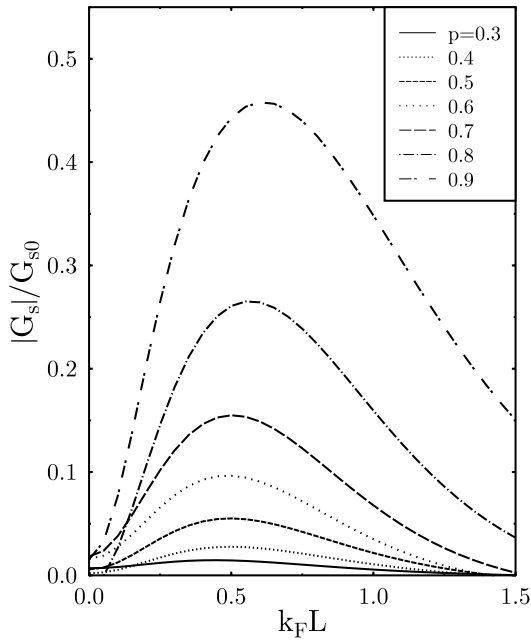}
\caption{Relative spin conductance of the wire with a domain wall
calculated as a function of its width $L$ for indicated values of the
parameter $p$.}
\end{figure}

In a nonmagnetic case we have $v_\uparrow = v_\downarrow$ and
therefore $G_s=0$. In the case considered here, $G_s=0$ when
there is no DW. Let us introduce the spin conductance for one
(spin-up) channel only, $G_{s0}=e/2\pi \hbar $. The relative spin
conductance in the presence of DW, $G_{s}/G_{s0}$, calculated
using Eq.~(19) and with numerically found transmission
coefficients, is shown in Fig.~3 as a function of the DW width
$L$ and for the indicated values of the parameter $p$. It
corresponds to the spin current outside the region of the domain
wall. The spin current inside the wall is not conserved because
of the spin-flip transitions.

In accordance with Eqs.~(19), (20) and (6), the nonzero spin current
in a one-channel wire with domain wall is due to a
difference in spin-flip transmissions for spin-up and spin-down
channels: the corresponding transmission coefficient turns out to
be larger for faster (majority) electrons.

\section{Spin polarization due to domain wall}

Spin dependent reflections from the wall lead to additional spin
polarization of the system near the wall. The distribution of
spin density created by the wall can be calculated using the
basis of scattering states. The $z$-component of the spin
density in the equilibrium situation ($U=0$) is
\begin{equation}
\label{24}
S_z(z)=-i\sum _n\int \frac{dk }{2\pi} \int
\frac{d\varepsilon }{2\pi}\; {\rm e}^{i\varepsilon \eta}\,
G_n(k,\varepsilon )\; \chi _n^\dag (z)\, \sigma _z\, \chi _n(z).
\end{equation}
The above formula  contains a constant part corresponding to the
spin density in the absence of DW, as well as the $z$-dependent
part $\delta S_z(z)$ created by the wall,
\begin{eqnarray}
\label{25}
\delta S_z(z) =\frac1{\pi }\int _0^{k_{F\uparrow }}dk\;
r_{R\uparrow }\, \cos(2k_\uparrow z)\hskip2cm
\nonumber \\
-\frac1{\pi }\int
_0^{k_{F\downarrow }}dk\; r_{R\downarrow }\, \cos(2k_\downarrow
z), \hskip0.5cm (z<-L),
\nonumber \\
=\frac1{\pi }\int _0^{k_{F\uparrow }}dk\; r_{L\uparrow }\,
\cos(2k_\uparrow z)\hskip2cm
\nonumber \\
-\frac1{\pi }\int _0^{k_{F\downarrow }}dk\;
r_{L\downarrow }\, \cos(2k_\downarrow z),
\hskip0.5cm (z>L).
\end{eqnarray}

\begin{figure}
\includegraphics[scale=0.45]{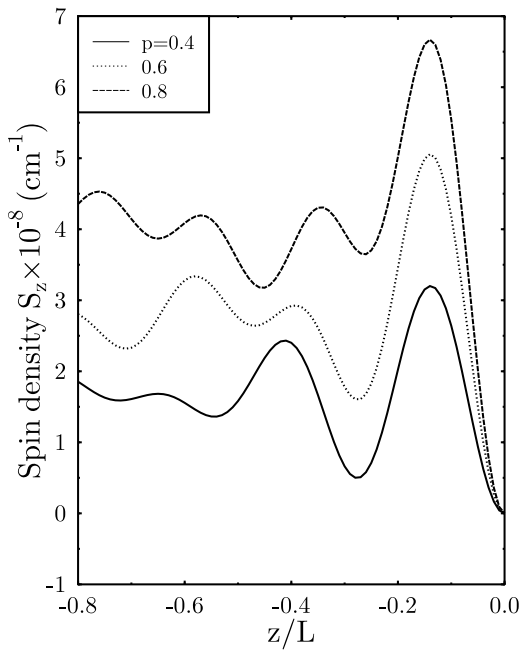}
\caption{Distribution of the spin density of electron gas near a domain
wall for different values of $p$.}
\end{figure}

The dependence of the spin density $S_z$ on the distance from the
wall is shown in Fig.~4. The spin dependent reflections from the
wall create spatial oscillations of the electron spin density.
These oscillations are similar to the Friedel oscillations of
charge in a nonmagnetic metal. However, one should point out here
that in addition to the above calculated spin polarization, there
is also a nonequilibrium spin polarization due to flowing
current.\cite{ebels00}

\section{Summary and concluding remarks}

We have presented in this paper a theoretical description of the
resistance of a magnetic microjunction with a constrained domain
wall at the contact. In the limit of $k_{F\uparrow (\downarrow )
}L\ll 1$, the electron transport across the wall was treated
effectively as electron tunneling through a spin-dependent
potential barrier. For such narrow and constrained domain walls
the electron spin does not follow adiabatically the magnetization
direction, but its orientation is rather fixed. However, the
domain wall produces some mixing of the spin channels.

The calculations carried out in the paper were restricted to a
limiting case of a single quantum transport channel. Accordingly,
the system was described by a one-dimensional model. However,
such a simple model turned out to describe qualitatively rather
well the basic physics related to electronic transport through
constrained domain walls, although the magnetoresistance obtained
is still smaller than in some experiments. In realistic
situations one should use a more general model. When the domain
wall does not cause transition between different channels, then
the description presented here can be applied directly to the
multichannel case by simply adding contributions from different
channels.

A domain wall leads to spin dependent scattering of conduction
electrons. Therefore, it also leads to a net spin polarization at
the wall, which oscillates with the distance from the wall,
similarly to Friedel oscillations of charge density near a
nonmagnetic defect in a nonmagnetic metal. We have calculated the
equilibrium component of this spin polarization.

It should be also pointed out that our description neglects
electron-electron interaction. Such an interaction is known to be
important in one-dimensional systems, particularly in the limit
of zero bias. The interaction may lead to some modifications of
the results in a very small vicinity of $U=0$, but we believe
that the main features of the magnetoresistance will not be
drastically changed.

\begin{acknowledgments}
We thank P. Bruno for very useful discussions. This work is
partly supported by Polish State Committee for Scientific Research
through the Grant No.~PBZ/KBN/044/P03/2001 and INTAS Grant No.~2000-0476.
\end{acknowledgments}

\end{document}